\documentclass[a4paper,12pt]{article}

\usepackage{citesort}
\usepackage{epsfig}
\usepackage{amsmath,amsfonts,amssymb}

\makeatletter

\@addtoreset{equation}{section}
\makeatother
\def\lsim{\;\raise0.3ex\hbox{$<$\kern-0.75em\raise-1.1ex\hbox{$\sim$}}\;}
\def\gsim{\;\raise0.3ex\hbox{$>$\kern-0.75em\raise-1.1ex\hbox{$\sim$}}\;}
\def\bit{\begin{itemize}}    \def\eit{\end{itemize}}
\def\beq{\begin{equation}}   \def\eeq{\end{equation}}
\def\noi{\noindent}
\def\tanb{\tan\beta}
\def\l{\lambda}
\def\k{\kappa}
\def\mueff{\mu_{\rm eff}}
\def\Beff{B_{\rm eff}}
\def\susy{{\sc susy}}
\def\Msusy{M_{\rm susy}}
\def\MGUT{M_{\rm GUT}}
\def\NMHDECAY{{\sf NMHDECAY}}
\def\NMSPEC{{\sf NMSPEC}}
\def\NMSSMTools{{\sf NMSSMTools\_1.0}}
\def\micromegas{{\sf micrOMEGAs\_2.0}}

\begin{document}

\centerline{\Large\bf NMSPEC: A Fortran code for the sparticle}
\vskip 3 truemm
\centerline{\Large\bf and Higgs masses in the NMSSM with}
\vskip 3 truemm
\centerline{\Large\bf GUT scale boundary conditions}
\vskip 2 truecm

\begin{center}
{\bf Ulrich Ellwanger}\\
\vskip 5 truemm
Laboratoire de Physique Th\'eorique\footnote{Unit\'e mixte de 
Recherche -- CNRS -- UMR 8627} \\
Universit\'e de Paris XI, F-91405 Orsay Cedex, France\\
\vskip 5 truemm
{\bf Cyril Hugonie}\\
\vskip 5 truemm
Laboratoire Physique Th\'eorique et Astroparticules\footnote{Unit\'e mixte de 
Recherche -- CNRS -- UMR 5207}\\
Universit\'e de Montpellier II, F-34095 Montpellier, France
\vskip 2 truecm

\end{center}

\begin{abstract} 
\NMSPEC\ is a Fortran code that computes the sparticle and Higgs masses,
as well as Higgs decay widths and couplings in the NMSSM, with soft
\susy\ breaking terms specified at $\MGUT$. Exceptions are the soft
singlet mass $m_s^2$ and the singlet self coupling $\kappa$, that are
both determined in terms of the other parameters through the
minimization equations of the Higgs potential. We present a first
analysis of the NMSSM parameter space with universal \susy\ breaking
terms at $\MGUT$ -- except for $m_s$ and $A_\kappa$ -- that passes
present experimental constraints on sparticle and Higgs masses. We
discuss in some detail a region in parameter space where a SM-like
Higgs boson decays dominantly into two CP odd singlet-like Higgs
states.
\end{abstract}

\vskip 1 truecm

PAC numbers: 12.60.Jv, 14.80.Cp, 14.80.Ly

\vfill
\noi December 2006\\
\noi LPT Orsay 06-79\\
\noi LPTA Montpellier 06-61

\newpage 
\pagestyle{plain} 
\baselineskip 18pt

\section{Introduction}

The Next to Minimal Supersymmetric Standard Model
(NMSSM)~\cite{nmssm,nmssmnum}
provides a very elegant solution to the $\mu$ problem of the MSSM  via
the introduction of a singlet superfield ${S}$. For the
simplest possible scale invariant form of the superpotential, the
scalar component of ${S}$ acquires naturally a vacuum
expectation value of  the order of the \susy\ breaking scale, giving
rise to a value of $\mu$  of order the electroweak scale. Hence the
NMSSM is the simplest supersymmetric extension of the standard model in
which the fundamental Lagrangian contains just \susy\ breaking terms
but no other parameters of the order of the electroweak scale.

As in the MSSM, the phenomenology of the NMSSM depends on a certain
number of parameters (mostly soft \susy\ breaking parameters) that
cannot be predicted from an underlying theory at present. It is then
useful to have computer codes that compute physically relevant
quantities as Higgs and sparticle masses, couplings, decay widths etc.
as functions of the initial parameters in the Lagrangian. Such codes
allow to investigate which regions in parameter space are in conflict
with present constraints on physics beyond the standard model and, most
importantly, which regions in parameter space can be tested in future
expe\-riments and/or astrophysical measurements. A review over
corresponding publicly available computer codes can be found in~\cite{bsm}.

Most of these codes are limited to the MSSM; the only presently
available code for the NMSSM is \NMHDECAY~\cite{nmhdecay1,nmhdecay2}. 

Many mechanisms for spontaneous \susy\ breaking (such as supergravity
with universal kinetic terms for all matter fields) predict a simple
pattern for the soft \susy\ breaking parameters at a large scale as the
GUT scale. Then, the soft \susy\ breaking parameters are often
universal, i.e. completely specified by universal gaugino masses,
scalar masses squared and trilinear couplings. In the case of the MSSM,
the corresponding scenario is denoted as the CMSSM (or mSUGRA since, in
principle, other scenarios as GMSB or AMSB could constrain the soft
\susy\ breaking parameters). The advantage of such a scenario is, apart
from its motivation through a simple underlying theory, obviously an
enormous reduction of the number of unknown parameters. It would be
desirable, if one could implement such simple boundary conditions also
in the NMSSM, which we will denote as the CNMSSM.

Numerical studies of the sparticle and Higgs spectrum in the CNMSSM
have been performed in~\cite{nmssmnum}. Since then, additional
radiative corrections to the Higgs spectrum have been computed~\cite{yuk},
and experimental constraints on the sparticle and notably
the Higgs spectrum have become considerably stronger~\cite{lhw}. Also,
the corresponding numerical codes have not been made publicly
available. 

\NMSPEC\ is a code that allows to chose parameters for the NMSSM at the
GUT scale. Exceptions are the singlet self coupling $\k$ and the 
soft \susy\ breaking singlet mass $m_s^2$; as  described below, these
two parameters are determined in terms of others through the
minimization equations of the Higgs potential. Possible scenarios such
as GMSB or AMSB are not (yet) implemented in \NMSPEC. Also, like
\NMHDECAY, \NMSPEC\ is at present limited to the simplest version of the
NMSSM; neither terms linear in the singlet~\cite{tad}, additional
U(1) gauge symmetries~\cite{adu1} or combinations thereof~\cite{gnmssm} 
are considered. (See~\cite{blls} for a review of such more general
versions of the NMSSM.) Subsequently we limit
ourselves to a scale invariant superpotential of the form
\beq
W = \l  {S}  {H}_u  {H}_d + \frac{\k}{3} \,  {S}^3 + \dots\ , 
\label{supot}
\eeq
and the associated soft \susy\ breaking trilinear couplings
$A_\l$, $A_\k$ and Higgs masses.

In principle numerical codes, that are designed to compute the particle
and sparticle spectrum depending on the soft \susy\ breaking terms
defined at the GUT scale, could proceed as follows: starting with
numerical values for all unknown parameters at $\MGUT$, the RG
equations can be integrated numerically down to the weak scale. Then,
the effective Higgs potential can be minimized numerically, and from the
Higgs vevs at the minimum one obtains $\tanb$ and $M_Z$ (where
$M_Z$ is used to determine a previously unspecified overall mass
scale). This approach has been used for the NMSSM in~\cite{nmssmnum}.

It has several disadvantages, however: First, the numerical minimization
of the Higgs potential is quite computer time consuming. Second, since
the overall mass scale and $\tanb$ are determined only at the end,
one finds that $m_{top}$, Higgs and sparticle masses often violate
experimental bounds -- it is not possible to input ``large'' soft terms
(and the correct value of $h_t$) from the start.

An advantage of this approach would be that it makes required (fine
tuned) relations among the initial parameters immediately obvious.

However, in view of its disadvantages, practically all present numerical
codes (within the MSSM) proceed differently: both $M_Z$ and $\tanb$
are used as inputs, and the two minimization equations of the Higgs
potential w.r.t. the two real Higgs vevs $H_u$ and $H_d$ are used to
compute $\mu$ and the soft $B$ parameter in terms of the others. (This
leaves open the sign of $\mu$; subsequently the two possibilities have
to be treated separately.) $\mu$ and $B$ have only a small effect on the
RG evolution of the other parameters, only via threshold effects from
particles whose masses depend on $\mu$ and/or $B$. One can take care of
this via a (typically rapidly converging) iterative procedure. (Now,
however, possible fine tunings originating notably from the requirement
to chose $B$ very precisely, are hidden.)

At first sight, an application of this procedure to the NMSSM is not
obvious: both $\mu$ and $B$ are no longer independent parameters
(although effective $S$ dependent parameters $\mueff=\l S$,
$\Beff = A_\l + \k S$ can still be defined), and one has to
cope with three coupled minimization equations w.r.t. $H_u$, $H_d$ and
$S$.

A possible way out is the following: First, the tree level minimization
equations w.r.t. $H_u$ and $H_d$ can be solved for $\mueff$ and
$\Beff$, as in the MSSM, in terms of the other parameters (incl. $M_Z$
and $\tanb$). From $\mueff$ and $\Beff$ one can deduce (for
$\l$ and $A_\l$ given) both $S$ and $\k$. Finally,
from the minimization equation w.r.t. $S$, one can easily obtain the
soft singlet mass $m_s^2$ in terms of all other parameters.

This allows to chose as input parameters in the Higgs sector of the
NMSSM the soft Higgs doublet masses squared, $A_\l$, $A_\k$,
$M_Z$, $\tanb$ and $\l$. From these the three parameters
$\mueff$ (or $S$), $\k$ and $m_s^2$ can be derived. The
radiative corrections to the Higgs potential (that depend on all
sparticle and Higgs masses) show a weak dependence on these parameters.
They can be included in the minimization equations that become
non-linear in the parameters to solve for, but one can solve them
iteratively by a loop that converges rapidly.

The derived parameters $\k$ and $m_s^2$ affect the RG evolution
equations of some of the other parameters not only through threshold
effects around $\Msusy$, but through the $\beta$ functions (already
at one loop). However, the numerical impact is relatively small such
that an iterative procedure converges quite rapidly again. (These
iterative procedures are described in section~2.)

Clearly, once $m_s^2$ is an output rather than an input, it becomes
difficult to find parameters such that $m_s^2$ at the GUT scale assumes
the same value as, e.g., the Higgs doublet (or other scalar) soft
masses squared. On the other hand it is easy to imagine that a mecanism
for the generation of soft \susy\ breaking terms treats the singlet
differently from the other non-singlet matter multiplets.

It happens frequently that the derived value of $m_s^2$ is negative
even at the GUT scale. This does not imply that the singlet vev takes
values of the order of the GUT scale, or that the scalar potential is
unbounded from below: due to the term $\k^2 S^4$ in the scalar
potential, the singlet vev will always be of the order of $m_s/\k$
(hence of the order of $\Msusy$), and it is still necessary to
integrate the RG equations also for the couplings and masses involving
the singlet down to the \susy\ scale and to compute further radiative
corrections there.

Once $\k$ is an output rather than an input, it is difficult to
study the ``Peccei-Quinn'' limit $\k \to 0$ (with $\l$
finite) in the NMSSM~\cite{mnz}. However, this limit is quite
unphysical at least in the simplest version of the NMSSM considered
here: Once $\k$ tends to zero, the singlet vev $S$ tends to
infinity, and hence $\mueff = \l S$ becomes unacceptably
large: large $\mueff$ implies a stable Higgs doublet potential at
the origin (hence no electroweak symmetry breaking), unless at least
one of the Higgs doublet \susy\ breaking masses squared is of the same
order and negative. Hence $\k \to 0$ would require $\Msusy \to
\infty$.

The limit where $\k$ and $\l$ tend to zero simultaneously and
the NMSSM turns into the MSSM plus a decoupled singlet sector poses no
problem, however: it is easy to see that, for given and fixed
$\mueff$, $\Beff$ and $A_\l$, $\l \to 0$ implies
automatically $\k \sim \l \to 0$. This limit is also stable
under the RG flow from the GUT scale down to the electroweak scale.
Although the vev $S$ tends towards infinity in this limit, the masses
of all components of the singlet superfield remain of the order of
$\Msusy$ (i.e. of the order of $A_\k$, $A_\l$ and
$\mueff$). Hence, in order to study the MSSM limit of the NMSSM
within \NMSPEC, it is enough to chose a tiny value for $\l$ on
input. ($\l = 0$ is not allowed, since expressions proportional to
$\k/\l$ would become ill defined. Also, $A_\k$ still has
to be chosen within a $\mueff$ and $A_\l$ dependent window in
order to avoid negative masses squared for the CP even and/or the CP
odd singlet scalars.)

Unless modified by the user, \NMSPEC\ allows to chose the following
input  parameters: $\tanb$ and the sign of $\mueff$ at the weak
scale and $\l$ at the \susy\ scale; at the GUT scale the free
parameters are universal gaugino masses $M_{1/2}$, universal scalar
masses $m_0^2$ and universal trilinear couplings $A_0$. Exceptions are
the following:\\ 
a) the soft singlet mass at $\MGUT$ is an output, as described above;
\\ b) the
trilinear coupling $A_\k$ can be chosen differently from the other
trilinear couplings. The reason for this is twofold: first, if an
underlying mecanism for the generation of the soft \susy\ breaking terms
treats the singlet differently from the other matter fields (as it is
already assumed for $m_s^2$), this will also affect the coupling
$A_\k$ which involves singlets only. (We left $A_\l$ unified
with the other trilinear couplings at $\MGUT$; this could easily be
changed by the user, however.) 

Also, interesting physics is associated with particular values of
$A_\k$: in certain regions of the parameter space of the NMSSM,
the lightest neutral CP even doublet-like Higgs boson (the one
ressembling to the SM Higgs boson) can decay into two lighter neutral
CP odd (sometimes also CP even) Higgs bosons~\cite{lighta,finet}. This
process allows for a doublet-like CP even Higgs boson $h$ to escape
LEP constraints even for $m_h$ below $114$~GeV. (The corresponding LEP
constraints have recently been updated in~\cite{lhw}, see the
discussion in section 3 below. The allowed window has become somewhat
tighter, but still exists.) On the one hand, such scenarios can make
the detection of a CP even Higgs boson at the LHC quite difficult~\cite{lighta,finet}.
It has also been argued that the fine tuning,
that is required in the MSSM in order to lift the Higgs mass above the
LEP bound, is relieved in the corresponding region of parameters in
the NMSSM~\cite{finet}. This region of parameters in the NMSSM
requires, however, that $A_\k$ assumes values within a certain
range such that singlet-like Higgs bosons are sufficiently light. The
possibility to chose $A_\k$ differently from the other trilinear
couplings in \NMSPEC\ allows to study this range (now in terms of
$A_\k$ defined at $\MGUT$), and the physics associated with it.

In the next section 2 we describe in detail, how the code \NMSPEC\ deals
with the various input parameters in the NMSSM. In section 3 we discuss
new features of some of the subroutines of \NMHDECAY\ {\sf v3}; most of the
subroutines of \NMHDECAY\ {\sf v3} are shared by \NMSPEC. In section 4 we
present first results obtained with \NMSPEC. The allowed parameter space
(passing all phenomenological tests) divides into three regions: a
MSSM-like region with $\l$ relatively small, and two NMSSM specific
regions
with $\l$ large either for small $\tanb$ and $m_h$ above 114~GeV,
or for medium to large $\tanb$ and $m_h$ below 114~GeV, but $A_\k$
within a certain window that renders the unconventional Higgs decay
modes discussed above possible. The width of the corresponding window
(and the fine tuning associated with it) is discussed in some detail.

\section{The code NMSPEC}

As discussed in the introduction, the code \NMSPEC\ proceeds as follows:
Input parameters are the SM gauge couplings, Fermion masses of the 3rd
generation, $\tanb$, the sign of $\mueff$, and $\l$ as well
as the soft \susy\ breaking masses and trilinear couplings (with the
exception of $m_s^2$) at the GUT scale.

The code \NMSPEC\ proceeds as follows: First, a guess for $\k$,
$m_s^2$, $\mueff$ and $\Msusy$ is made, and a preliminary
calculation of $\MGUT$ and the gauge and Yukawa couplings at the GUT
scale is performed through an integration of the (two loop) RG
equations from the \susy\ scale up to $\MGUT$ using the subroutine
RGES. ($\MGUT$ is defined as the scale where $\alpha_1$ and
$\alpha_2$ unify.)

Then, an ``external'' loop starts, that involves both the GUT scale and
$\Msusy$: the (two loop) RG equations for the soft terms, together
with the gauge and Yukawa couplings, are integrated from $\MGUT$ down
to $\Msusy$ in the subroutine RGESINV.

There an ``internal'' loop starts, that involves physics between
$\Msusy$ and the weak scale:

In the subroutine RUNPAR, first the mass $M_A^2$ of a potentially heavy
Higgs doublet is estimated (which is required for the threshold effects
below). The \susy\ scale (denoted by $Q2$) is defined in terms of the
squark masses of the first generations, and a scale QSTSB is defined in
terms of the left and right stop masses squared. Then, the SM gauge and
Yukawa couplings are calculated at the scale QSTSB (together with the
Higgs wave function renormalization constants $Z_i$). Threshold effects
at Higgs and sparticle masses are taken care of. Finally, the NMSSM
specific Yukawa and trilinear couplings are computed at QSTSB.

In the next subroutine MSFERM, the ($\mueff$-dependend) 3rd
generation squark and slepton pole masses and mixing angles are computed
at the scale QSTSB. They are needed below for the radiative corrections
to the minimization equations of the Higgs potential.

In the following subroutine LOWMUK, first the RG equations for the soft
Higgs doublet masses squared are integrated from $Q2$ to QSTSB. Then,
$\mueff$, $\k$ and $m_s^2$ are determined from the minimization
equations of the Higgs potential including radiative corrections, that
depend on previous estimates for $\mueff$ and $\k$. A variable
CHECK verifies the relative change of $\mueff$ and $\k$ with
respect to its previous values. Finally, an improved value for $M_A^2$
is computed (in case the internal loop stops here; otherwise its
previous estimate is used again).

Unless CHECK $< 10^{-6}$, the internal loop returns now to the
subroutine RUNPAR. There, the improved values for $\mueff$ and 
$\k$ are used for improved threshold corrections. In the next call
of LOWMUK, the radiative corrections to the minimization equations
also include the improved values of $\mueff$ and $\k$.

If CHECK $< 10^{-6}$, $\mueff$ and $\k$ have been determined to
a sufficient accuracy. We have never observed a convergence problem
related to the determination of $\mueff$ and $\k$; 5 inner loops
are usually more than sufficient.

Then, the ``external'' loop continues: next, the RG equations for the
SM gauge and Yukawa couplings are integrated again from the weak scale
up to $\MGUT$ in the subroutine RGES, using improved threshold
effects at sparticle and Higgs masses as well as the updated value for
$\k$. The value for $\MGUT$ is improved simultaneously.

Next, the RG equations for the soft terms are integrated from
$\Msusy$ up to $\MGUT$ in the subroutine RGESUNI, using the new SM
gauge and Yukawa couplings
and the new values for $\MGUT$ and $m_s^2$. A variable GUTEST
verifies, whether all soft terms coincide with the required boundary
conditions at $\MGUT$.

Unless the relative deviation determined by GUTEST is less than
$10^{-4}$ (for all soft \susy\ breaking parameters), the external loop
jumps back to a call of RGESINV. There, the new values for $\MGUT$
and the SM gauge and Yukawa couplings are used in order to integrate
again the RG equations for the soft terms (with the required boundary
conditions at $\MGUT$) down from the GUT scale.

If GUTEST is less than $10^{-4}$, the required values of the soft terms
at $\MGUT$ are considered as sufficiently precise, and the external
loop is left. The subsequent subroutines of \NMSPEC\ are shared with
\NMHDECAY\ {\sf v3}: Higgs, gluino, chargino and neutralino masses are
computed, and all Higgs branching ratios into SM particles and
sparticles are determined. Then, present experimental constraints on
Higgs and sparticles masses and couplings are verified. Warnings
indicate at the end, whether such constraints are
violated. A rough (1 loop) computation of the $b \to s\ \gamma$
branching ratio is performed (for information only, without generating
an error message), and it is verified whether the choice of parameters
corresponds to a global minimum of the NMSSM Higgs potential.

Finally, if desired (if the flag OMGFLAG is on), the dark matter relic
density is computed using a NMSSM version of the \micromegas\
subroutines~\cite{micro}. These subroutines are called internally in
\NMHDECAY\ {\sf v3} and \NMSPEC; an interface file, 
as it was the case for \NMHDECAY\ {\sf v2}~\cite{microld},
is not needed anymore. The NMSSM
version of \micromegas\ is included in the downloadable package
\NMSSMTools, and compiled together with \NMHDECAY\ and
\NMSPEC, as described in the appendix.

The different options (a single point in parameter space, scans over
regions in parameter space using grids or random numbers) are also
described in the appendix.

\section{New features in NMHDECAY v3 and NMSPEC} 

The code \NMHDECAY\ {\sf v3} performs the same tasks as the codes \NMHDECAY\
{\sf v1} and {\sf v2}, that are described in detail in~\cite{nmhdecay1,nmhdecay2}.
Now, many subroutines are shared by \NMSPEC, and the codes \NMHDECAY\ {\sf v3}
and \NMSPEC\ can be downloaded simultaneously (see the appendix). Apart
from the architecture, some properties of some subroutines have been
improved with respect to \NMHDECAY\ {\sf v2}; these changes are described
below.

\subsection{Experimental constraints} 

The (negative) results of the four LEP collaborations ALEPH, DELPHI, L3
and OPAL on Higgs Boson searches have been updated and combined by the
LEP Higgs Working group in~\cite{lhw}. As compared to the corresponding
constraints already implemented in \NMHDECAY\ {\sf v2}, the updates in~\cite{lhw}
include improved constraints on Higgs masses vs. couplings
for the following processes:
\bit
\item $Z \to Z H_2 \to Z H_1 H_1$ where $H_1$ stands for a CP even or a
CP odd neutral Higgs boson. The considered decay channels of $H_1$ are
$H_1 \to 2b$ and $H_1 \to 2\tau$, and combinations thereof.

\item $Z \to H_1 H_2$ where both $H_1$ and $H_2$ can denote a CP even
or a CP odd neutral Higgs boson. Decays of both $H_1$ and $H_2$ into
$2b$ and $2\tau$ are studied.

\item $Z \to H_1 H_2 \to H_1 H_1 H_1$. Only simultaneous decays of all
three $H_1$ states either into $6b$ or into $6\tau$ have been
considered.
\eit

The corresponding
data files are included in the directory EXPCON. They restrict the
allowed parameter space of the NMSSM more strongly than
within the previous versions {\sf v1} and {\sf v2} of \NMHDECAY.
In particular, the constraints on the decay of a SM-like Higgs boson
$h$ to $aa$, where $a$ is a CP odd (singlet-like) scalar, are much
stronger. Roughly speaking, the only remaining possibilities (unless
the  $Z$-$Z$-$h$ coupling is heavily suppressed) are:
\bit
\item
86~GeV $\lsim m_h \lsim$ 106~GeV and $m_a \lsim$ 11~GeV (where
the $b\bar{b}$ decay of $a$ is impossible, and no constraints from
$a \to \tau\tau$ are available),
\item
106~GeV $\lsim m_h \lsim$ 114~GeV with $m_a$ going up to $m_h/2$.
\eit
(It must be noted, however, that LEP constraints are implemented only for
individual processes. Combinations of different processes can rule out
points in parameter space, that seem to have passed the individual
constraints.)

\subsection{Input parameters} 

The implicit definitions of the input parameters $\l$, $\k$,
$\mueff$, $A_\l$ and $A_\k$ in \NMHDECAY\ {\sf v3} differ slightly
from the ones in \NMHDECAY\ {\sf v2}: now these are all defined at a common
SUSY scale $Q2$ (essentially the average squark masses of the first two
generations, unless $Q2$ is defined by the user), in the 
${\overline{DR}}$ scheme as all other soft \susy\ breaking parameters.
The numerical effect concerns essentially $A_\l$: in order to
reproduce data points obtained with the previous versions of \NMHDECAY,
$A_\l$ has to be shifted by an $A_{top}$ dependend amount (that
can be estimated from the RG equation for $A_\l$).

\subsection{Higgs mass} 

The calculation of the radiative corrections to the Higgs masses has
been reorganised and improved: Now all parameters used by the
corresponding subroutine MHIGGS (gauge and Yukawa couplings,
$\tanb$, trilinear couplings, quark and squark masses, $\mueff$
and Higgs vevs) are defined at a scale QSTSB, that corresponds to an
average of the right and left stop quark masses. (QSTSB is computed
separately from the scale $Q2$ above, and can be lower.) Corresponding
formulas for the radiative corrections have essentially become simpler,
except that contributions from gauge boson and light Higgs boson loops
have now to be added explicitly. (In the LLA, which is used for the
elektroweak corrections, contributions from particles with masses $\sim
M_Z$ to masses of particles of the same order do not involve large
logarithms. Now, the results on heavy Higgs masses are more reliable,
without negative effects on the precision of the light Higgs Boson
mass.)

Inside the subroutine MHIGGS, dominant gaugino loop contributions are
now expressed in the form of a shift of $A_\l$, which also
improves the precision of heavy Higgs spectrum.

New radiative corrections, that have not been considered before, are
two loop contributions to the lightest Higgs mass involving two powers
of large logarithms and electroweak gauge couplings $g_{1/2}$ of the
form $\sim h_t^4\ g_{1/2}^2 \log^2$, where $\log$ denotes a potentially
large logarithm. The dominant effect of these contributions depends on
the chargino/neutralino spectrum, and is described in~\cite{2lnmssm}.

\section{The allowed parameter space of the CNMSSM} 

In the CMSSM, the present experimental constraints on sparticle and
Higgs masses can be satisfied by chosing sufficiently large soft \susy\
breaking terms at the GUT scale, and $\tanb$ not too small. As
stated above, the required fine tuning is hidden through the implicit
determination of $\mu$ and $B$.

In the CNMSSM (without unification of the soft singlet mass, see above)
it is often necessary, in addition, to chose a non-universal value for
$A_\k$ in order obtain positive masses squared for both CP even and
CP odd singlet scalars, even when they are essentially decoupled.
Otherwise, for $\l \lsim 0.15$, the allowed parameter space of the
CNMSSM corresponds essentially to the one of the CMSSM
(except for the possibility to have $m_h$ below 114~GeV and
unconventional Higgs decay modes, see below).

Larger values for $\l$ are not always allowed within the CNMSSM,
however; possible problems arise typically in the CP even Higgs sector:
on the one hand, non-vanishing $\l$ increases the mass of the
lightest Higgs doublet, at tree level, by an amount
\beq
\Delta m_h^2 = 2\l^2 M_Z^2 \sin^2 \! 2\beta /(g_1^2+g_2^2)\ .
\eeq 
This increase is relevant, however, only for small $\tanb$. On the
other hand non-vanishing $\l$ induces mixings between the doublet
and the singlet states. Since the singlet state is typically quite
heavy, this mixing reduces the mass of the lightest eigenstate of the
$3\times 3$ CP even mass matrix. This lightest eigenstate will then
often violate bounds from LEP~\cite{lhw}, in spite of its singlet
component.

In order to maximize the mass of the lightest CP even mass eigenstate,
this singlet-doublet mixing has to vanish (as described in~\cite{2lnmssm}),
which implies a relation between $\mueff$,
$A_\l$, $\l$, $\tanb$ and $\k$. This relation is
generally not satisfied within the CNMSSM; then -- at least for larger
values of $\tanb$ -- this mixing effect disallows often large
values of $\l$.

For given soft \susy\ breaking terms at the GUT scale, one then obtains
typically allowed ``islands'' in the plane $\l$ vs. $\tanb$.
An example is given in the Figure 1 below, which shows such an allowed
island for $M_{1/2}= m_0=500$~GeV, $A_0=-800$~GeV, $A_\k=-1500$~GeV
and $\mueff >0$ (with $m_{top}=175$~GeV).

\begin{figure}[tb]
\begin{center}
\epsfig{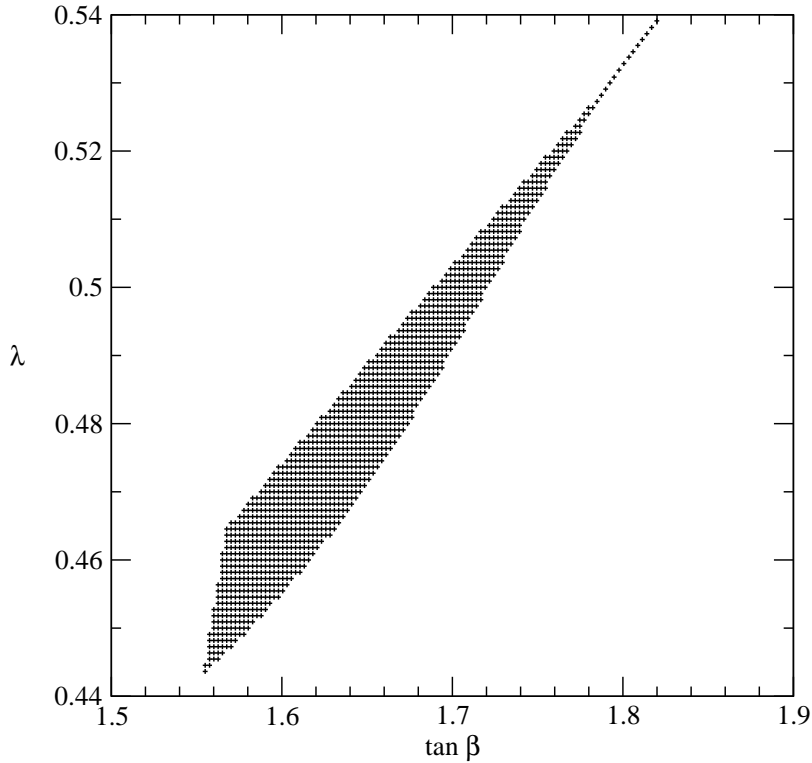}
\end{center}
\caption{Allowed values for $\l$ vs. $\tanb$ 
for $M_{1/2}= m_0=500$~GeV, $A_0=-800$~GeV, $A_\k=-1500$~GeV.}
\end{figure}

The lightest (doublet-like) Higgs mass is $\sim 114$~GeV near both ends
of this region, and up to $\sim 117.5$~GeV near its center. All three
Yukawa couplings $\l$, $\k$ and $h_t$ are relatively large, and
the region is bounded -- apart from LEP constraints on $m_h$ -- by the
condition on the absence of a Landau singularity in the Yukawa sector
below $\MGUT$.

Larger values of $\l$ are also possible for larger values of
$\tanb$. Then, however, the lightest (still doublet-like) Higgs
mass falls below $114$~GeV, and $A_\k$ has to be chosen within a
relatively narrow window in order to generate a light (singlet-like) CP
odd scalar into which $h$ can decay~\cite{lighta,finet}. Such a line of
allowed values for $A_\k$ vs. $\l$ is shown in Figure 2, for
$\tanb=5$, $M_{1/2}= m_0=500$~GeV and $A_0=-800$~GeV.

\begin{figure}[tb]
\begin{center}
\epsfig{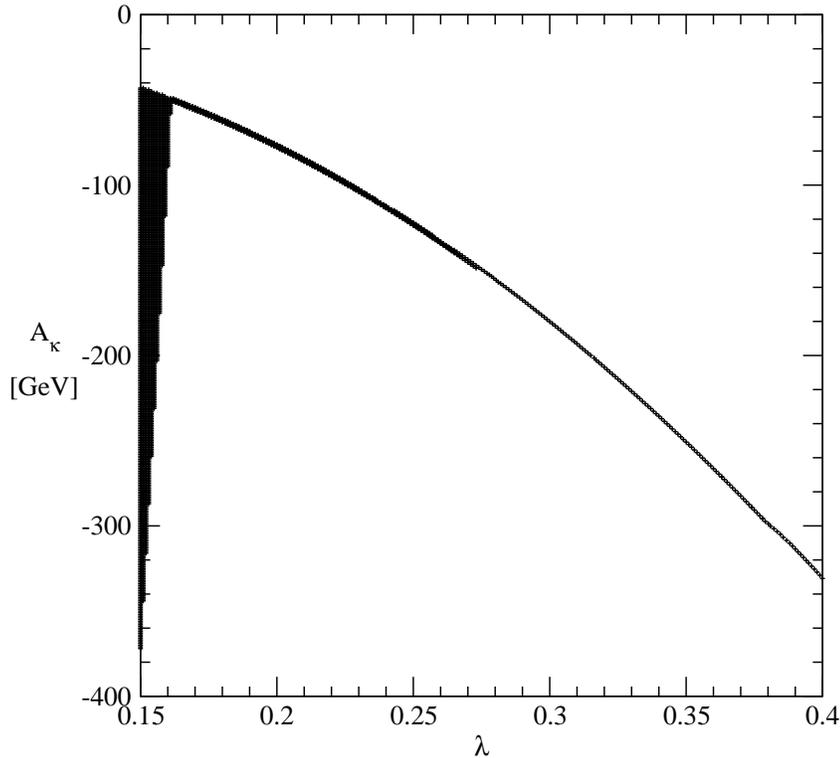}
\end{center}
\caption{Allowed values for $A_\k$ vs. $\l$ for $\tanb=5$, 
$M_{1/2}= m_0=500$~GeV and $A_0=-800$~GeV.}
\end{figure}

For $\l \lsim 0.16$, this line joins the CMSSM-like region where
$m_h > 114$~GeV. Along the line, $m_h$ decreases to $m_h \sim 86$~GeV
for $\l \sim .4$. Larger values of $\l$, i.e. lower
values of $m_h$, are excluded by LEP.
Across the line, the mass of the singlet-like CP odd scalar goes up to
$m_h/2$ for $\l \lsim .27$ (where $m_h \gsim$ 106 GeV), but $m_a \lsim
11$~GeV for larger values of $\l$ (where $m_h \lsim$ 106 GeV); a larger
value for $m_a$ would imply that $a$ decays mainly into  $b\bar{b}$,
which is excluded by LEP.
The width of the line, expressed in terms of $A_\k$, is $\sim 1.7$~GeV
for $\l \lsim .27$, but shrinks down to $\sim 100$~MeV for larger values of $\l$.
This allows to estimate the fine tuning in $A_\k$
required for this region of the parameter space of the CNMSSM.

In principle, the width of the line (and hence the required relative
fine tuning) can depend on all other soft \susy\ breaking terms. In
order to see this, we have chosen smaller values for $M_{1/2}= 200$~GeV,
$m_0 =100$~GeV, fixed $\l = 0.2$ (still at $\tanb=5$), and
plotted the allowed values for $A_\k$ vs. $A_0$ in Figure 3.

\begin{figure}[tb]
\begin{center}
\epsfig{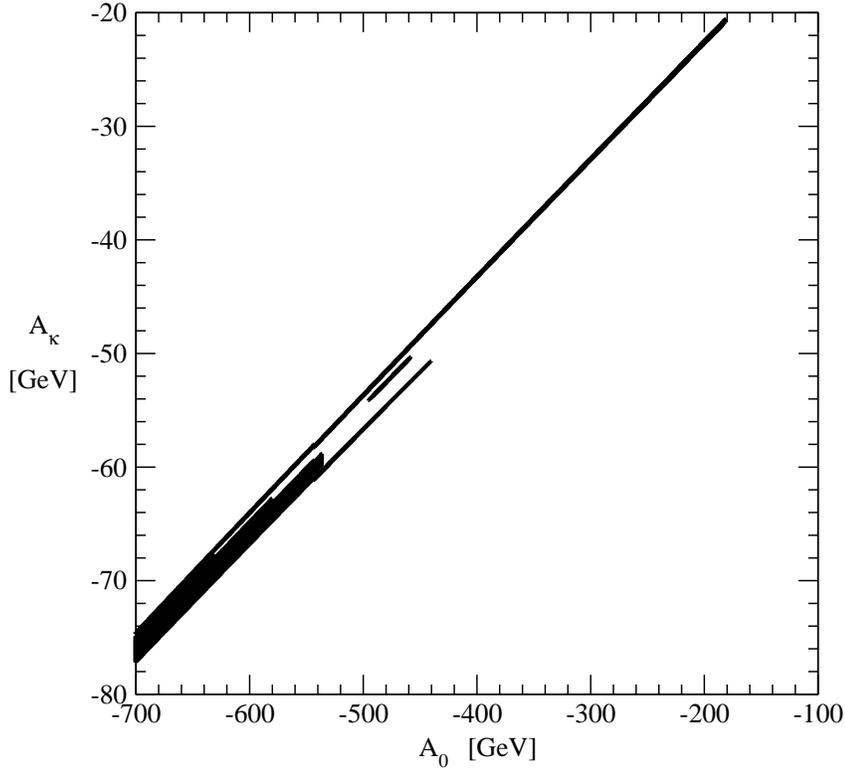}
\end{center}
\caption{Allowed values for $A_\k$ vs. $A_0$ for $\tanb=5$, 
$M_{1/2}= 200$~GeV, $m_0=100$~GeV and $\l=0.2$.}
\end{figure}

In Figure 3, $m_h \sim 90$~GeV near $A_0 \sim -200$~GeV and increases
up to $\sim 112$~GeV for $A_0 \sim -700$~GeV. (For values of $A_0 \lsim
700$~GeV one finds $m_{\widetilde t_1} \lsim 100$~GeV, which is
excluded by stop searches at LEP). 
The mass of the singlet-like CP odd scalar is again $\lsim 11$~GeV for
$A_0 \gsim -440$~GeV, but varies from a few GeV up to $m_h/2$ for $A_\k \lsim
-440$~GeV. The width of the line, expressed in terms of $A_\k$, is now
$\sim 2.5$~GeV for  $A_0 \sim -700$~GeV, i.e. the relative fine tuning
in $A_\k$ is somewhat less than before. According to our preliminary analysis,
this width cannot be increased by chosing different values of $\tanb$.

On the other hand allowed regions in parameter space with
$m_h < 114$~GeV remain present also for smaller values of $\lambda$ (at
least down to $\lambda \sim 0.1$).
The present analysis of the allowed parameter space of the CNMSSM is
far from complete, but \NMSPEC\ will allow for more detailed studies in
this direction in the future.
In any case it is important to note that the region in the parameter
space of the NMSSM where the lightest SM-like Higgs boson decays
dominantly into two CP odd scalars remains present in the CNMSSM; we
recall that it could then be quite challenging to detect just one Higgs
boson at the LHC~\cite{lighta}.

\newpage
\noi {\Large{\bf Acknowledgement}}
\vskip 2mm

We thank P. Bechtle for providing us with the data files containing
present constraints on Higgs masses and couplings as obtained by
the LEP Higgs Working group. We also thank A. Pukhov for help
concerning the development of a simple and global compilation procedure,
including all \micromegas\ subroutines. We acknowledge
support by the ANR grant PHYS@COL\&COS.

\newpage
\noi {\Large{\bf Appendix: How to use NMSSMTools}}
\vskip .7cm

The package \NMSSMTools\ contains \NMSPEC, the new version
of \NMHDECAY\ ({\sf v3}), a NMSSM version of
\micromegas\ and an updated version of the LEP constraints.
It can be downloaded from the web page:
{\sf http://www.th.u-psud.fr/NMHDECAY/nmssmtools.html}.
Once this compressed tar file is downloaded, one has to type
\begin{verbatim}
tar -zxvf NMSSMTools_1.tgz
\end{verbatim}
\noi which will create the directory \NMSSMTools. Inside this
directory one finds:

\bit
\item the directory {\sf sources}, that contains the subroutines common
to \NMHDECAY\ and \linebreak
\NMSPEC, the directory {\sf  micromegas\_2.0.3} with the subroutines for
the computation of the dark matter relic density, and the directory
{\sf EXPCON} with data files used for the experimental constraints;

\item the main programs for \NMHDECAY\ and \NMSPEC:
{\sf nmhdecay.f, nmhdecay\_rand.f,} \linebreak
{\sf nmspec\_grid.f, nmspec.f, nmspec\_rand.f and nmspec\_grid.f;}
 
\item six sample input files: {\sf inp.dat,  randinp.dat, gridinp.dat, inpsp.dat,
randinpsp.dat} and \linebreak
{\sf gridinpsp.dat}. All input (and output) files are in SLHA format~\cite{slha},
including some NMSSM specific switches that are described
in~\cite{nmhdecay1,nmhdecay2,slha2}. We have added privately defined
switches for the boundary values of the parameters to be scanned over,
the number of points to be scanned over, and an integer ISEED that
serves as input for the random number generator. A sample input file
{\sf randinpsp.dat} is given in Table 1;

\item corresponding output files: {\sf spectr.dat, decay.dat, omega.out,
randerr.dat, randout.dat,}
\linebreak {\sf griderr.dat, gridout.dat, spectrsp.dat, decaysp.dat,
omegasp.dat, outsp.dat, randerrsp.dat,}
\linebreak {\sf randoutsp.dat, griderrsp.dat, gridoutsp.dat;}

\item A master {\sf Makefile}.
\eit

\noi In order to compile the subroutines in the directories {\sf sources} and
{\sf  micromegas\_2.0.3}, one has to type first:
\begin{verbatim}
make init
\end{verbatim}
\noi Then to compile the main programs, type:
\begin{verbatim}
make
\end{verbatim}
\noi This will create 6 executable files: {\sf nmhdecay, nmhdecay\_rand,
nmhdecay\_grid, nmspec,} \linebreak 
{\sf nmspec\_rand} and {\sf nmspec\_grid}. Now the user has the following options:

\bit
\item To study a single point in parameter space, with input
parameters at the SUSY scale:
Edit the input file {\sf inp.dat}. Type {\sf nmhdecay}. The output
spectrum is in the file {\sf spectr.dat}, the branching ratios of the
6 Higgs states in the NMSSM are in the file {\sf decay.dat}. The
dark matter relic density (if OMGFLAG=1 in {\sf inp.dat}) is in the file
{\sf omega.dat}. (A first call of the computation of the dark matter relic
density provokes the compilation of additional subroutines of
\micromegas.)

\item To scan over randomly chosen points within specified
boundary values, with input parameters at the SUSY scale:
Edit the input file {\sf randinp.dat}. Type {\sf nmhdecay\_rand}. The
number of points that have passed all tests and their ranges are in the
file {\sf randerr.dat}. The output is in the file {\sf randout.dat}. In order to
specify the output, you need to edit the file {\sf nmhdecay\_rand.f}
which contains the subroutine {\sf OUTPUT}. At the beginning of this
subroutine, you find the different elements of the array {\sf RES},
whose elements are printed into the output file {\sf randout.dat}. You
can declare the elements of the array {\sf RES} according to your
needs; note that the input parameters are elements of the array {\sf
PAR} whose meaning is given at the beginning of the {\sf MAIN} program
in {\sf nmhdecay\_rand.f}. After saving a modified version of {\sf
nmhdecay\_rand.f} it is necessary to type {\sf make} again.

\item To scan over a lattice of points within specified
boundary values, with input parameters at the SUSY scale:
Edit the input file {\sf gridinp.dat}. Type {\sf nmhdecay\_grid}. The
number of points that have passed all tests and their ranges are in the file
{\sf griderr.dat}. The output is in the file {\sf gridout.dat}. In order to
specify the output, you need to edit the subroutine {\sf OUTPUT} in the
file {\sf nmhdecay\_grid.f}, and proceed as described in the case of a
random scan above.

\item To study a single point in parameter space, with input
parameters at the GUT scale:
Edit the input file {\sf inpsp.dat}. Type {\sf nmspec}. The output
spectrum is in the file {\sf spectrsp.dat}, the branching ratios of
the 6 Higgs states in the NMSSM are in the file {\sf decaysp.dat}. The
dark matter relic density (if OMGFLAG=1 in {\sf inp.dat}) is in the file
{\sf omegasp.dat}. The \susy\ scale parameters are written in the file
{\sf outsp.dat} in the same format as the file {\sf inp.dat}, so that
{\sf outsp.dat} can be used as an input file for {\sf nmhdecay}.

\item To scan over randomly chosen points within specified
boundary values, with input parameters at the GUT scale:
Edit the input file {\sf randinpsp.dat} (cf table~\ref{table1}).
Type {\sf nmspec\_rand}. The
number of points that have passed all tests and their ranges are in the
file {\sf randerrsp.dat}. The output is in the file {\sf randoutsp.dat}. In order
to specify the output, you need to edit the subroutine {\sf OUTPUT} in the file
{\sf nmspec\_rand.f}, and proceed as described above.

\item To scan over a lattice of points within specified
boundary values, with input parameters at the GUT scale:
Edit the input file {\sf gridinpsp.dat}. Type {\sf nmspec\_grid}. The number of
points that have passed all tests and their ranges are in the file {\sf griderrsp.dat}.
The output is in the file {\sf gridoutsp.dat}. In order to specify the output, you need
to edit the subroutine {\sf OUTPUT} in the file {\sf nmspec\_grid.f}, and proceed as
described above.
\eit

\noi To delete all the already compiled codes, the user should type:
\begin{verbatim}
make clean
\end{verbatim}

\begin{table}[p]
{\footnotesize
\baselineskip 10pt
\begin{verbatim}
# INPUT FILE FOR NMSPEC VERSION 1
# BASED ON SUSY LES HOUCHES ACCORD II

BLOCK MODSEL
  3     1               # NMSSM PARTICLE CONTENT

BLOCK SMINPUTS
  1   127.92D0          # ALPHA_EM^-1(MZ)
  2   1.16639D-5        # GF
  3   0.1172D0          # ALPHA_S(MZ)
  4   91.187D0          # MZ
  5   4.214D0           # MB(MB), RUNNING B QUARK MASS
  6   175.D0            # TOP QUARK POLE MASS
  7   1.777D0           # MTAU

BLOCK MINPAR
  0     0.              # REN.SCALE
  4     1.              # SIGMU
  17    100.            # M0MIN
  18    100.            # M0MAX
  27    200.            # M12MIN
  28    200.            # M12MAX
  37    5.0             # TBMIN
  38    5.0             # TBMAX
  57    -700.           # A0MIN
  58    -550.           # A0MAX

BLOCK EXTPAR
  617   .2D0            # LMIN
  618   .2D0            # LMAX
  647  -80.             # A_KMIN
  648  -55.             # A_KMAX
  
BLOCK STEPS
   0    1000             # NTOT
   1    -1              # ISEED
\end{verbatim}}
\caption{The {\sf randinpsp.dat} file for sample parameter scan.
\label{table1}}
\end{table}

\end{document}